\documentclass[11pt,preprint,graphicx]{aastex}
\usepackage{graphicx}

\def\etal{\it et al. \rm }

\begin{document} 

\title{Stellar Populations and the Star Formation Histories of LSB Galaxies:
II. HII Regions}

\author{James Schombert}
\affil{Department of Physics, University of Oregon, Eugene, OR 97403;
jschombe@uoregon.edu}

\author{Stacy McGaugh}
\affil{Department of Astronomy, Case Western Reserve University, Cleveland, OH 44106;
stacy.mcgaugh@case.edu}

\author{Tamela Maciel}
\affil{Department of Physics, Cambridge University, Cambridge, UK;
tm419@cam.ac.uk}

\begin{abstract}

\noindent The luminosities, colors and H$\alpha$ emission for 429 HII
regions in 54 LSB galaxies are presented.  While the number of HII regions
per galaxy is lower in LSB galaxies compared to star-forming irregulars and
spirals, there is no indication that the size or luminosity function of HII
regions differs from other galaxy types.  The lower number of HII regions
per galaxy is consistent with their lower total star formation rates.  The
fraction of total $L_{H\alpha}$ contributed by HII regions varies from 10
to 90\% in LSB galaxies (the rest of the H$\alpha$ emission being associated with
a diffuse component) with no correlation with galaxy stellar or gas
mass.  Bright HII regions have bluer colors, similar to the trend 
in spirals; their number and luminosities are consistent with the
hypothesis that they are produced by the same HII luminosity function as
spirals. Comparison with stellar population models indicates that the
brightest HII regions in LSB galaxies range in cluster mass from a few
$10^3 M_{\sun}$ (e.g., $\rho$ Oph) to globular cluster sized systems (e.g.,
30 Dor) and that their ages are consistent with clusters from 2 to
15 Myrs old.  The faintest HII regions are comparable to those in
the LMC powered by a single O or B star.  Thus, star formation in LSB
galaxies covers the full range of stellar cluster mass.

\end{abstract}

\section{Introduction}

The observational tracers of star formation range from the near-UV
(Boissier \etal 2008) to the far-IR (Bigiel \etal 2008) and, while each
wavelength region has its advantages and disadvantages, LSB galaxies are
difficult to observe outside the traditional optical bandpasses.  The most
visible feature of star formation in LSB galaxies is the H$\alpha$ line,
produced by young, massive stars that compose the upper end of the IMF.
The UV photons emitted by these stars will, in turn, ionize the surrounding
gas to form a HII region.  While the total H$\alpha$ luminosity of a galaxy
measures its global star formation history, these HII regions map the
amount and location of local star formation, providing a window in to the
details of the star formation process.

Studying HII regions in galaxies allows one to 1) investigate star
formation both globally and locally, 2) examine the upper mass limit of
stellar mass function and 3) map the structure of the ISM.  While UV and
far-IR emission may provide a more nuanced view of total star formation;
the size, location and luminosity of HII regions displays the local
variation of star formation directly and can be used to resolve stellar
population questions.  In addition, the size and luminosity of HII regions
provide information on number of ionizing stars and the mass of the
underlying stellar associations.

Previous work on HII regions in galaxies focused on high surface brightness
spirals and irregular galaxies (e.g., Caldwell \etal 1991, Kennicutt, Edgar
\& Hodge 1989, Youngblood \& Hunter 1999).  These studies found that
the number of HII regions in a galaxy increases with later Hubble type, in
correlation with the total star formation rate (SFR), and found various
differences in the HII luminosity function as a function of galaxy
properties.  However, very little work has been completed on H$\alpha$
emission in low surface brightness (LSB) galaxies due to the technical
difficulty in measuring narrow band fluxes for object so close to the
brightness of the night sky.  Studies by Schombert \etal (1992), McGaugh,
Schombert \& Bothun (1995) and recent work by Kim (2007) represent the
deepest H$\alpha$ studies in LSB galaxies.

The results from these previous works can be summarized that LSB galaxies
have 1) small regions of H$\alpha$ emission (assumed to be low in
luminosity, although this early data was not flux calibrated), 2) weakly
correlated with regions of enhanced surface brightness and 3) no coherent
patterns indicative of density wave scenarios.  Small and weak HII regions
are consistent with the low SFR's for LSB galaxies as a class of objects,
and agreed with the hypothesis that these galaxies are quiescent and
inhibited in their star formation histories.

This paper, the second in our series on optical observations of PSS-II LSB
galaxies, presents the H$\alpha$ spatial results which map the size,
location and luminosities of HII regions in our sample galaxies.  With this
information, our goal is to compare the style of star formation in LSB galaxies
with spirals and irregulars to detect any global differences in their star
formation histories.  The characteristics of importance to the star
formation history of a galaxy are the number of HII regions, the luminosity
of the brightest HII regions, the shape of the HII region luminosity
function and the spatial positions of HII regions with respect the optical
distribution of light.  Lastly, we examine the optical colors of the HII
regions in the hope of resolving the color dilemma in LSB galaxies, their
unusually blue colors, yet low total SFR's.

\section{Analysis}

Observations, reduction techniques and the characteristics of the sample
are described in Paper I (Schombert, Maciel \& McGaugh 2011).  Our final
sample contains 58 LSB galaxies selected from the PSS-II LSB catalog
(Schombert, Pildis \& Eder 1997) with deep $B$, $V$ and H$\alpha$ imaging
from the KPNO 2.1 meter.  H$\alpha$ emission was detected in 54 of the 58
galaxies.  All detected galaxies had at least one distinct HII region,
although diffuse emission account for approximately 50\% of the total
H$\alpha$ emission in most LSB galaxies.  

The sample galaxies all have irregular morphology with some suggestions of
a bulge and disk for a handful.  They range in size from 0.5 to 10 kpc and
central surface brightnesses from 22 to 24 mag arcsces$^{-2}$.  Their total
luminosities range from $-$14 to $-$19 $V$ mags, which maps into stellar
masses from 10$^7$ to 10$^9$ $M_{\sun}$.  The gas fractions for the sample
are between 0.5 and 0.9, so the amount of HI gas covers a similar range.

Identification of a HII region followed a slightly different prescription
from previous studies.  In our case, we have identified an H$\alpha$ knot
to be a HII region if it is 1) distinct, i.e. not a filament or diffuse
region, 2) has rough circular symmetry (where spatial resolution limits
this determination) 3) having a clear peak in H$\alpha$ emission, and 4)
falling off uniformly around the peak.  Due to resolution limits, any
particular region may include several HII complexes for more distant
galaxies in the sample.  However, even for the most distant galaxies, one
arcsec corresponds to 400pc which is sufficient to resolve the high
luminosity HII regions into smaller components.  There was no correlation
with the number of HII regions and distance (see \S4) which would imply
that confusion is not a factor in our sample.

Identification was made by visually guiding a threshold algorithm applied
to smoothed H$\alpha$ images.  The center of confirmed HII knots were
determined and the luminosity of each selected region was determined by a
circular aperture.  The radius of the aperture is determined to be the
point where the flux falls to 25\% of the peak emission.  This value is
used for the size of the HII region, regardless of any indication of
non-circularity.

\begin{figure}[!ht]
\centering
\includegraphics[scale=0.3,angle=0]{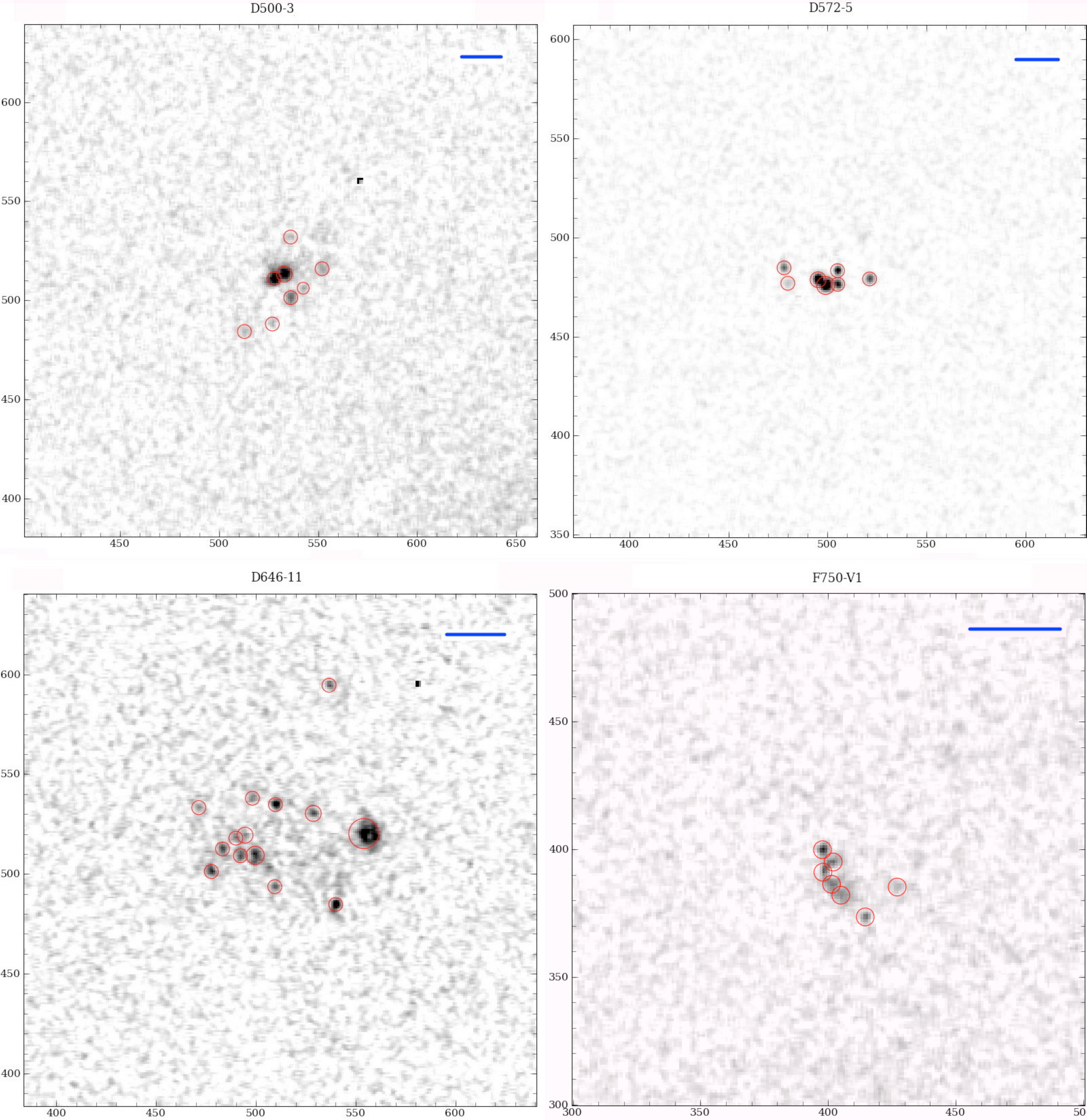}
\caption{\small H$\alpha$ maps for four galaxies in our sample.  The
selected HII regions are indicated using our criteria of distinctiveness and symmetry.
The blue bar indicates a spatial scale of 5 kpc.
}
\label{hii_apertures}
\end{figure}

Four examples of our HII region selection process is shown in Figure
\ref{hii_apertures} where the selected HII regions are shown inside red
circles.  A five kpc scale is indicated in each frame.
Continuum images (Johnson $V$) can be found at the our data
website (http://abyss.uoregon.edu/$\sim$js/lsb), as well as all the
information on individual HII regions plus color and surface brightness
data on the sample.  The four examples in Figure \ref{hii_apertures} were
selected to illustrate several key points about the HII regions in LSB
galaxies.

Galaxy D500-3 (upper left) displays two bright regions near the galaxy core
and a number of fainter regions surrounding the core.  None of the HII
regions are evident as higher continuum surface brightness regions from $V$
frames.  Even though the brightest two regions are relatively high in
H$\alpha$ luminosity (38.24 and 38.17 log L$_{H\alpha}$, approximately 20
Orion complexes), their stellar populations have no effect on the optical
structure of the nearby region of the galaxy.  The 5 kpc bar is indicated
in the upper right of the frame, where the larger HII regions are 100 to
150 pc in size, ranging down to 25 pc for the fainter regions.

Galaxy D572-5 (upper right) exhibits a more luminous set of HII regions
from other LSB galaxies, again several bright regions in the core and a few
fainter HII regions in the outer regions.  There is some indication of
diffuse H$\alpha$ emission in the outer disk, but insufficient to warrant
inclusion by our selection algorithm.  The brighter HII regions are just
visible in the continuum $V$ frames as distinct blue knots.

Galaxy D646-11 (lower left) displays more scattered H$\alpha$ emission.
The selected HII regions are not centrally concentrated.  In fact, the
brightest region (more of a shell or bubble than a star complex) is located
in the outer disk.  There are several filaments and diffuse H$\alpha$
regions in the core that were not selected as HII regions.  The brighter
HII regions are associated with bluer continuum colors, but this is not
always the case for LSB galaxies as a whole (Pildis, Schombert \& Eder 1997).

Galaxy F750-V1 (lower right) is a smaller, nearby LSB galaxy.  While seven HII
regions were selected, most of its H$\alpha$ emission is diffuse.  It is a
subjective determination to select any knot in the core region.  There is
no signature from the HII regions in the continuum images; however, there
is enhanced blue stellar colors in the diffuse regions.

Similar criteria to identify H$\alpha$ knots were used to identify surface
brightness knots in the $V$ frames.  The mean surface brightness isophotes
(based on ellipse fits) are subtracted from the raw image.  This subtracted
image is threshold searched for optical knots.  As with the H$\alpha$
knots, these regions are marked and measured with circular apertures
defined by the 25\% width.  In the final analysis, 492 HII regions were
identified in 54 LSB galaxies and two DDO objects (154 and 168).  In
addition, 271 optical knots were identified in the $V$ frames.  Of the
492 HII regions, 207 had no distinct optical counterpart.  Of the 271 $V$ knots,
only 49 had no detectable H$\alpha$ emission.  The properties of these
regions will be discussed in \S7.

\section{HII Regions Sizes and Luminosities}

\begin{figure}[!ht]
\centering
\includegraphics[scale=0.7,angle=0]{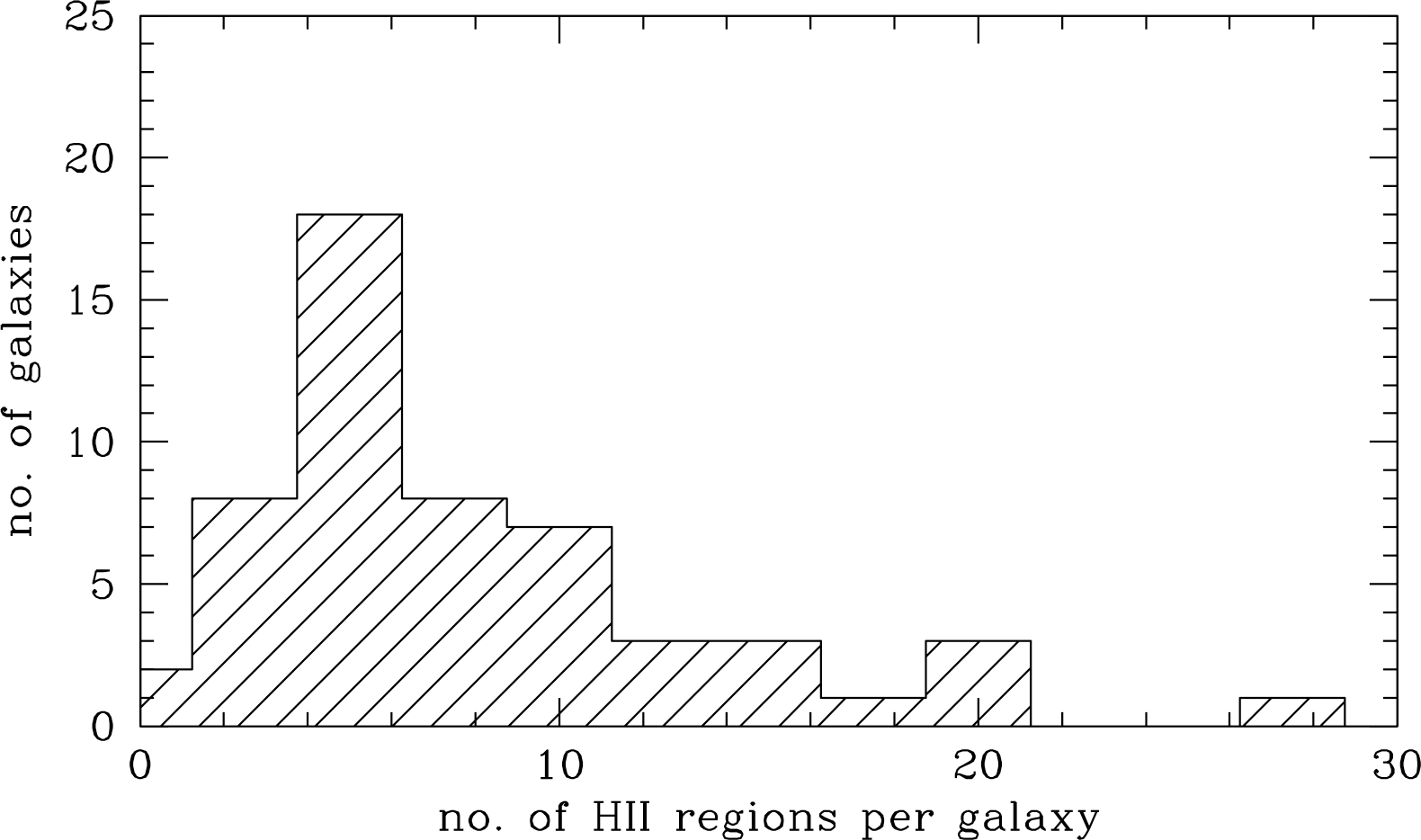}
\caption{\small A histogram of the number of HII regions found in each galaxy.  The
number we find, per galaxy, is typically much lower than other studies due
to our more stringent selection criteria, with most LSB galaxies having less
than 10 HII regions.  However, LSB galaxies still
display much lower numbers of HII regions than other star-forming galaxy
types, in line with their low total SFR's.
}
\label{num_hii_regions}
\end{figure}

In our total LSB sample, 54 (93\%) galaxies had more than one identifiable
HII region.  The four galaxies undetected by our H$\alpha$ imaging had the
four lowest gas fractions (less than 0.4).  A histogram of the number of
HII regions per galaxy is shown in Figure \ref{num_hii_regions}.  The 
typical of HII regions per galaxy is between 3 and 10, which is quite low
for late-type galaxies with irregular morphology (Caldwell \etal 1991) but
consistent with values from early studies of H$\alpha$ emission in LSB
galaxies (McGaugh, Schombert \& Bothun 1995).  We note that these mean
values are much less than the numbers found by Youngblood \& Hunter (1999)
for HII regions in dIrr's.  That number is usually above 20 HII regions per
galaxy; but, this is due in part to our different selection schemes and the
intrinsic nature of rich, star-forming dIrr's.  We have two galaxies in
common, DDO154 and DDO168.  Youngblood \& Hunter find 74 and 58 HII
regions, respectfully, whereas we only find 14 and 25 for the same systems.
While this might appear that we are incomplete in our HII region selection,
the total H$\alpha$ fluxes are in agreement and the difference in number
simply reflects our more stringent selection criteria in defining clear,
isolated HII regions, rather than H$\alpha$ filaments.

\begin{figure}[!ht]
\centering
\includegraphics[scale=0.8,angle=0]{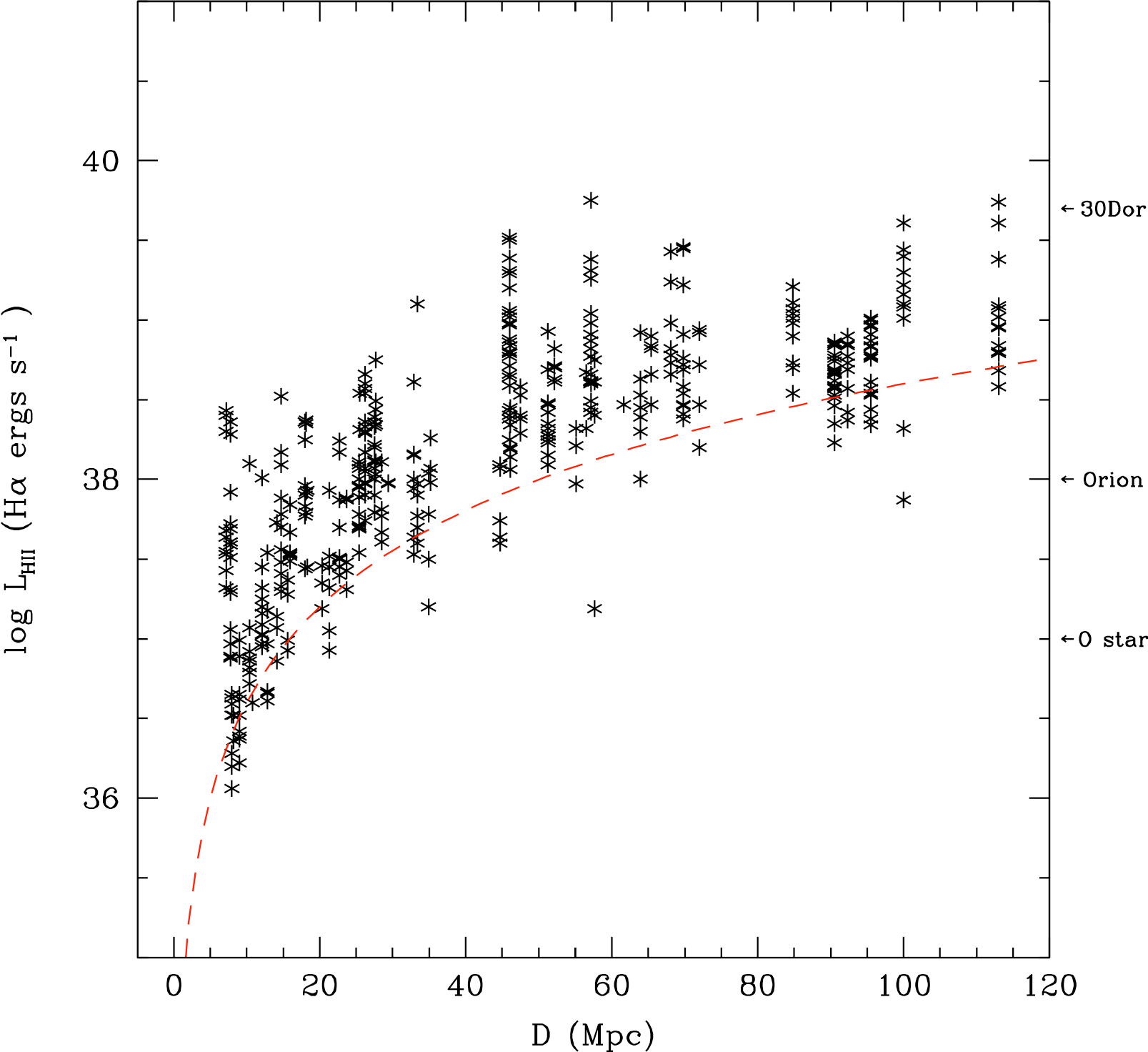}
\caption{\small The H$\alpha$ luminosity of individual HII regions
($L_{HII}$) as a
function of galaxy distance.  Fainter HII regions are missing from the
sample of galaxies farther than 40 Mpc due to decreasing spatial/luminosity resolution
(a $L \propto D^2$ cutoff is shown).  The brightest HII regions are found
in the more distant galaxies, indicating that 30 Doradus sized star forming
complexes are rare in LSB galaxies, and a larger volume of the Universe
must be searched to locate them.
}
\label{dist_flux}
\end{figure}

The H$\alpha$ luminosities for all the HII regions in our sample is shown
in Figure \ref{dist_flux} (note we distinguish the total H$\alpha$ of a
galaxy, $L_{H\alpha}$, versus the H$\alpha$ luminosity of an individual HII
region, $L_{HII}$).  The HII region luminosities range from $5 \times
10^{36}$ ergs s$^{-1}$ for the faintest regions to 10$^{39.5}$
for the brightest regions.  A single O7V star results in a HII region of
log $L_{HII} = 37.0$ (Werk \etal 2008), although HII regions powered by single B0
stars are found in the LMC with $L_{HII} = 36.0$ to 36.2 (Zastrow, Oey \&
Pellegrini 2013).  Thus, the faintest regions are difficult to explain
under the observation that very few O or B stars are born in isolation (Chu \&
Gruendl 2008) or may be the result of PN ionization (Walterbos \& Braun
1992).  The brighter regions correspond to a 30 Doradus sized complexes and
would contain $10^6 M_{\sun}$ solar masses of $H_2$ gas; however,
even these individual regions would not be detected in CO surveys of LSB
galaxies (Schombert \etal 1990).

In some ways, the distribution of HII region luminosities in LSB galaxies
are similar to the distribution in early-type spirals rather than irregulars.  In
early-type spirals, there are more low luminosity HII regions relative to
the brightest ones (Kennicutt, Edgar \& Hodge 1989), with fewer of the
massive star forming regions found in dwarf irregulars.  On the other hand,
LSB galaxies with HII regions brighter than log $L_{HII} > 38$ do exist, but
HII regions of this size are not found in Sa spirals (Caldwell \etal
1991).  Thus, it seems the HII regions in LSB galaxies follow more closely
the pattern of other galaxies with irregular morphologies; unfortunately,
we lack sufficient statistics to construct a HII luminosity function for
individual galaxies in order to rigorously examine this effect.

Flux completeness for our HII region selection is a greater concern for our
sample, for we explore a larger volume of the Universe than other samples
as the original PSS-II catalog was surface brightness selected with an
angular size limit, not luminosity limited.  The individual HII region
luminosities are shown in Figure \ref{dist_flux} as a function of galaxy
distance.  As can be seen in this Figure, the brightest HII regions are
found in the most distant galaxies (which are also the most
massive/brightest galaxies).  In addition, the galaxies farther than 40 Mpc
are deficient in HII regions fainter that log $L_{H\alpha}$ = 38.
Interestingly, the 40 Mpc limit is the same limited distance found by
Kennicutt, Edgar \& Hodge (1989) based on resolution experiments with their
H$\alpha$ imaging study.

The lack of fainter HII regions for the more distant galaxies is probably
due to a lack of spatial resolution to distinguish a HII complex from
diffuse H$\alpha$ emission.  To test this hypothesis, we selected a subset
of galaxies between 20 and 30 Mpc and deconvolved their H$\alpha$ images to
simulate their appearance at 80 to 120 Mpc.  As expected, the fainter HII
regions (log $L_{HII} < 38$) dropped below the threshold of detection.
However, due to the typical wide spacing of HII regions in LSB galaxies,
there was no significant increase in the brightness of the remaining HII
regions due to blending.  We conclude that our sample will severely under
sample low luminosity HII regions for objects greater than 40 Mpc in
distance and, thus, any discussion of a HII region luminosity function must
take this bias into account.

Even for the more complete nearby portion of our sample ($D < 40$ Mpc) the
ratio of $L_{HII}/L_{H\alpha}$ is dramatically different from those found
by Youngblood \& Hunter.  Their distribution (their Figure 10) displays
very few galaxies with ratios less than 80\%, such that a majority
of H$\alpha$ emission comes from distinct star forming regions, although
the determination method differs from our calculations in the sense that
they assign HII regions to complexes then compare the amount of H$\alpha$
flux from complexes versus their total fluxes.  For our sample, a
significant amount of H$\alpha$ emission in LSB galaxies (typically 50\%)
arises from a warm, diffuse component, rather than directly from HII
complexes, in agreement with the dwarf galaxies studied by van Zee (2000).
The ionizing source of this diffuse component is difficult to determine
(Hoopes \etal 2001).  Although this conclusion is strongly dependent on
whether one can isolate small, weak HII regions in the diffuse component,
objects that our more stringent selection criteria would miss.

Another concern is that the brightest HII regions are found in the most
distant galaxies.  This may be due to confusion, where the HII regions
selected by this study are, in fact, blends of fainter HII regions blurred
by distance.  While this may be true for some individual cases, the number
of HII regions as a function of distance does not show a decreasing trend
with distance, a relationship one would expect if a number of fainter HII
regions are being mistakenly grouped together as one complex.  The more
likely trend is that fainter HII regions are simply indistinct and confused
with diffuse H$\alpha$ emission, therefore, not selected by our criteria.

\begin{figure}[!ht]
\centering
\includegraphics[scale=0.8,angle=0]{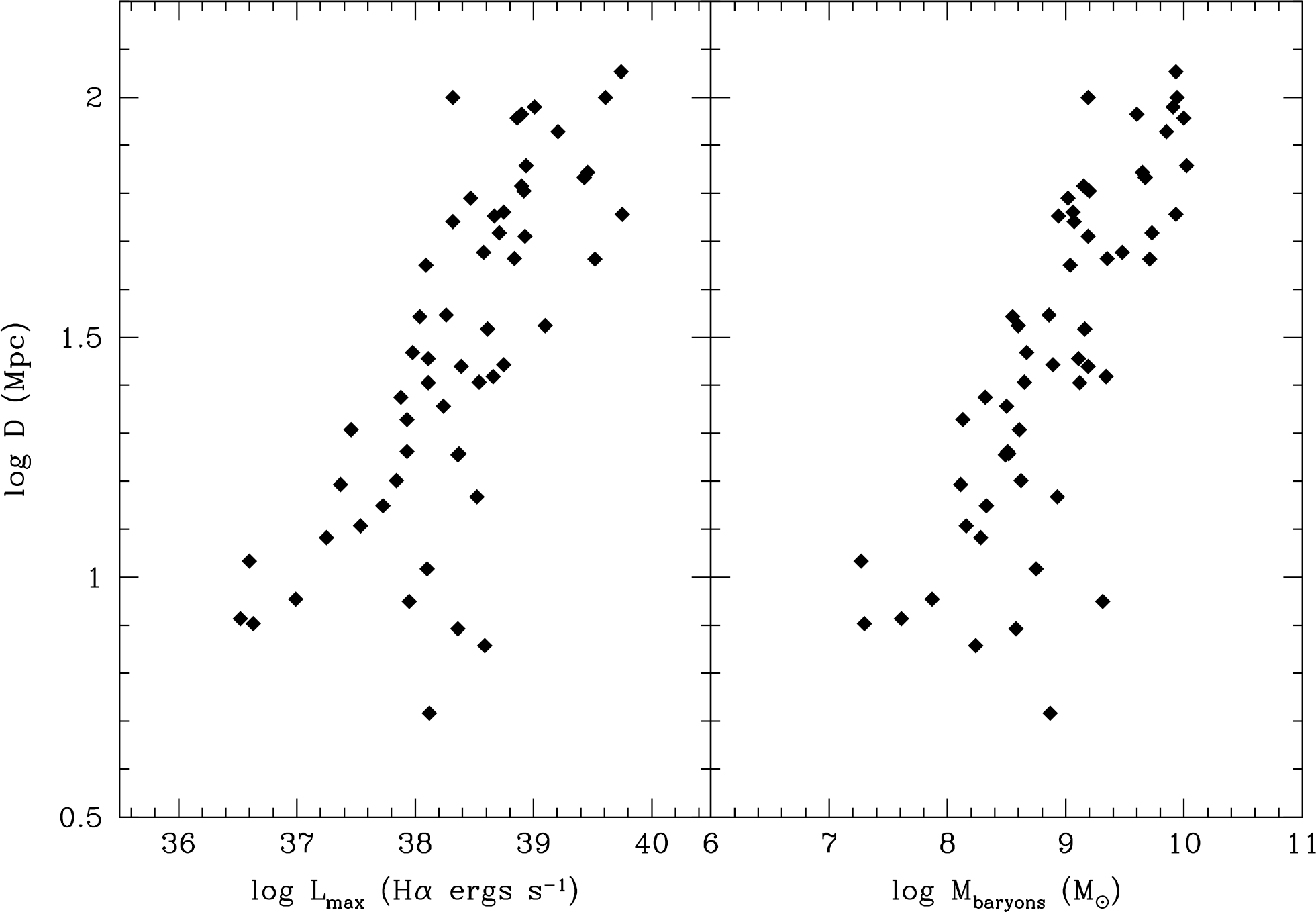}
\caption{\small The luminosity of the brightest HII region in a galaxy and
baryon mass as a function of distance.  The brightest HII regions and most
massive galaxies are also the most distant objects in our sample, i.e. a
larger volume of the Universe must be sample to find the largest LSB
galaxies.
}
\label{lmax_dist}
\end{figure}

Figure \ref{lmax_dist} displays the luminosity of the brightest HII region
($L_{max}$) and the baryon mass of a galaxy (stellar mass plus gas mass) as
a function of distance.  As noted in Paper I, the most massive LSB galaxies
in our sample are at the largest distances.  The brightness of the
brightest HII region also increases with distance, in synchronous with the
baryon mass (see \S 6).  We conclude that the reason that the brightest HII
regions in LSB galaxies are found in the most distant galaxies is due to a
volume selection effect.  The more distant objects in our sample are the
brightest by luminosity (and the largest in baryon mass) and are also
galaxies with the highest H$\alpha$ fluxes in the sample.  The low mass,
low H$\alpha$ luminosity galaxies in the sample would not be found at large
distances due to the angular size limit to the PSS-II catalog.  There is no
reason to believe that Malmquist bias plays a role in our sample, as it was
not selected by total or H$\alpha$ luminosity.  The brighter HII regions in
distant galaxies simply reflect the diversity of LSB galaxies, where LSB
galaxies with bright 30 Doradus sized star forming complexes are rare.
However, due to the loss of fainter HII regions with distance in the
sample, in our following discussions we will distinguish between the
distant sample ($D > 40$ Mpc) and the more complete nearby sample.

\begin{figure}[!ht]
\centering
\includegraphics[scale=0.8,angle=0]{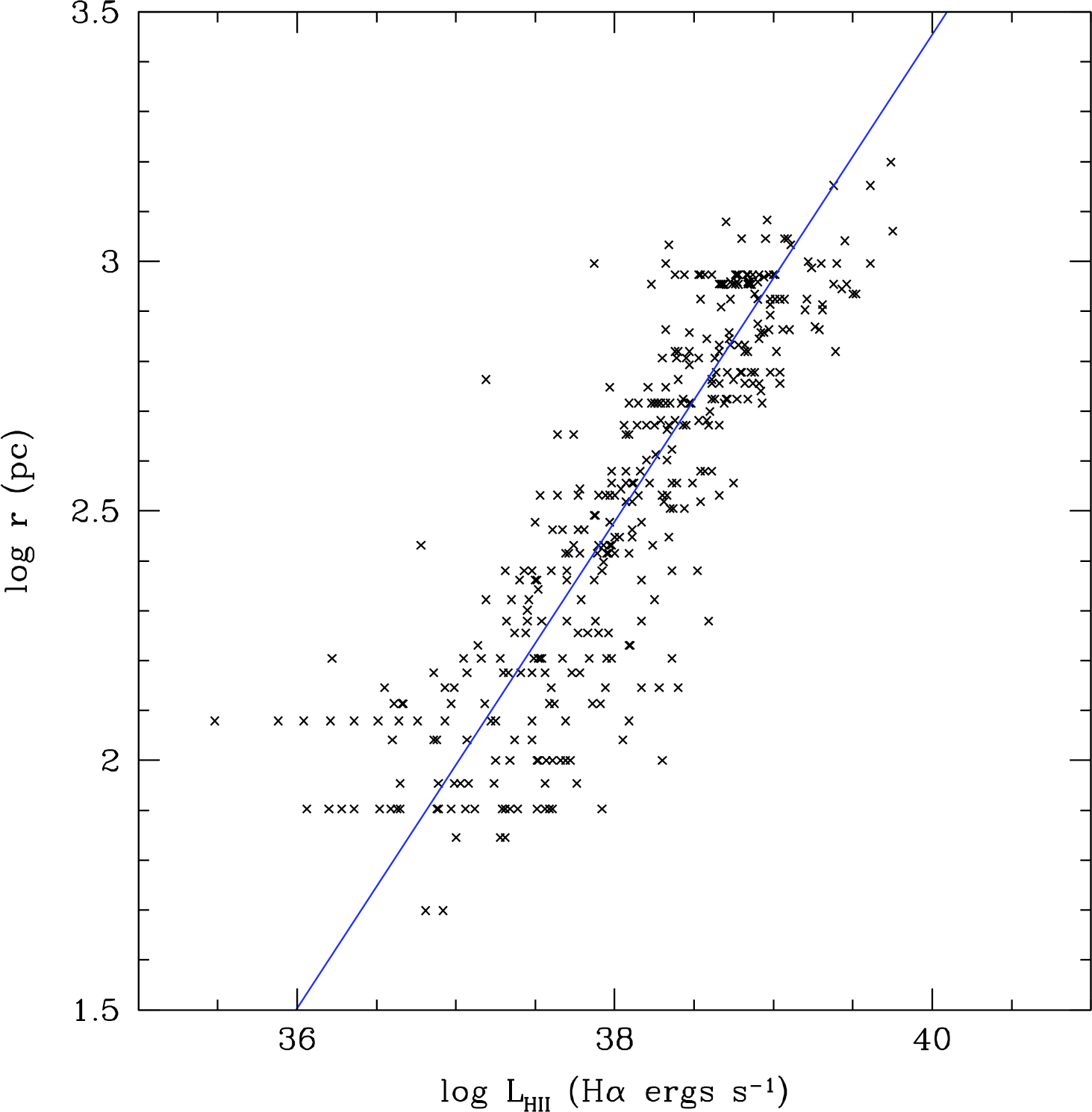}
\caption{\small The size of a HII region in parsecs versus the H$\alpha$
luminosity of the same region.  A linear fit (blue line) is consistent with
a relation of log $L_{H\alpha} \propto r^{2}$ meaning that there is little
extinction by dust in LSB galaxy HII regions.
}
\label{flux_radius}
\end{figure}

For the sample as a whole, about 50\% of the imaged galaxies have between
75 to 200 pc/pixel resolution, 25\% have a resolution less than 50
pc/pixel, where the radius of a HII region is estimated by the point where
the flux falls to 25\% the peak flux.  A plot of HII region radius ($r$, in
pc's) versus their H$\alpha$ luminosity is shown in Figure
\ref{flux_radius}.  The slope of the relationship is consistent with log
$L_{HII} \propto r^{2}$, meaning that we detect all the H$\alpha$ photons
produced in the complexes.  Foreground extinction by dust is very small in
LSB galaxies compared to spirals, in agreement with the lack of far-IR
detection for LSB galaxies and their low mean metallicities (Kuzio de
Naray, McGaugh \& de Blok 2004).  Hence, we make no corrections for
internal extinction in any of our quoted flux values.

\section{HII Region Numbers}

\begin{figure}[!ht]
\centering
\includegraphics[scale=0.8,angle=0]{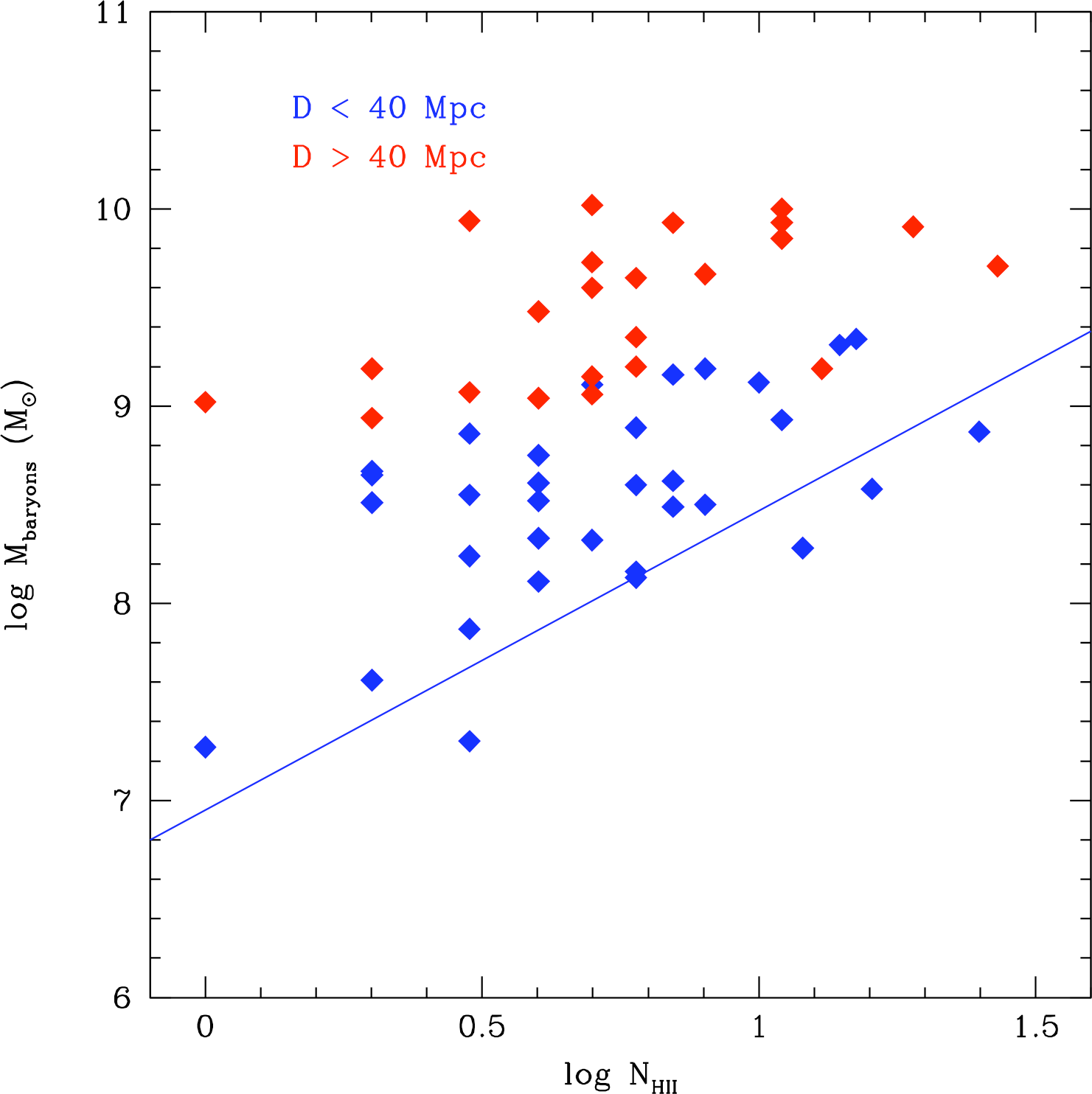}
\caption{\small The number of HII regions per galaxy as a function of
baryon mass (gas plus stellar mass, see McGaugh \& de Blok 1997).  The line
is from Youngblood \& Hunter (1999) for dwarf irregulars.  Our nearby
sample follows their relationship (albeit with lower total numbers due to
our more stringent selection criteria); however, the distant galaxies are
deficient in low luminosity HII regions.
}
\label{num_mass}
\end{figure}

The number of HII regions as a function of galaxy mass is shown in Figure
\ref{num_mass}.  There is a similar relation between number of HII regions
and galaxy mass as found by Youngblood \& Hunter (1999) (blue line in
Figure \ref{num_mass}).  Again, the distant galaxies in our sample fail to
display any relationship due to the under counting of fainter HII regions.
The nearby sample displays the same slope as Youngblood \& Hunter, although
our more stringent detection criteria shifts our number counts to lower
values.

The relationship between number and galaxy mass may simply reflect the
statistical effect of more gas material in a larger volume results in more
star formation events.  As star formation is driven by local density
(Helmboldt \etal 2005), then more volume will produce more individual star
forming regions.  There is a also a trend of brightest HII region flux with
the number of HII regions; but, again, this reflects the statistical
behavior of larger volume provides a greater chance of a larger star
formation event.

The number density of HII regions per kpc$^{-2}$ has a weak trend of
decreasing density with increasing galaxy mass where the typical number
density (for the $D < 40$ Mpc sample) is between 0.1 and 1 HII regions per
kpc$^{-2}$ with a mean of 0.3.  This is similar to the mean value for Sm/Im
type galaxies from Kennicutt, Edgar \& Hodge (1989).  There is no trend of
number density with galaxy mass/size; however, the $D < 40$ sample has a
limited dynamic range in galaxy size and mass.

\section{HII Regions Locations}

\begin{figure}[!ht]
\centering
\includegraphics[scale=0.8,angle=0]{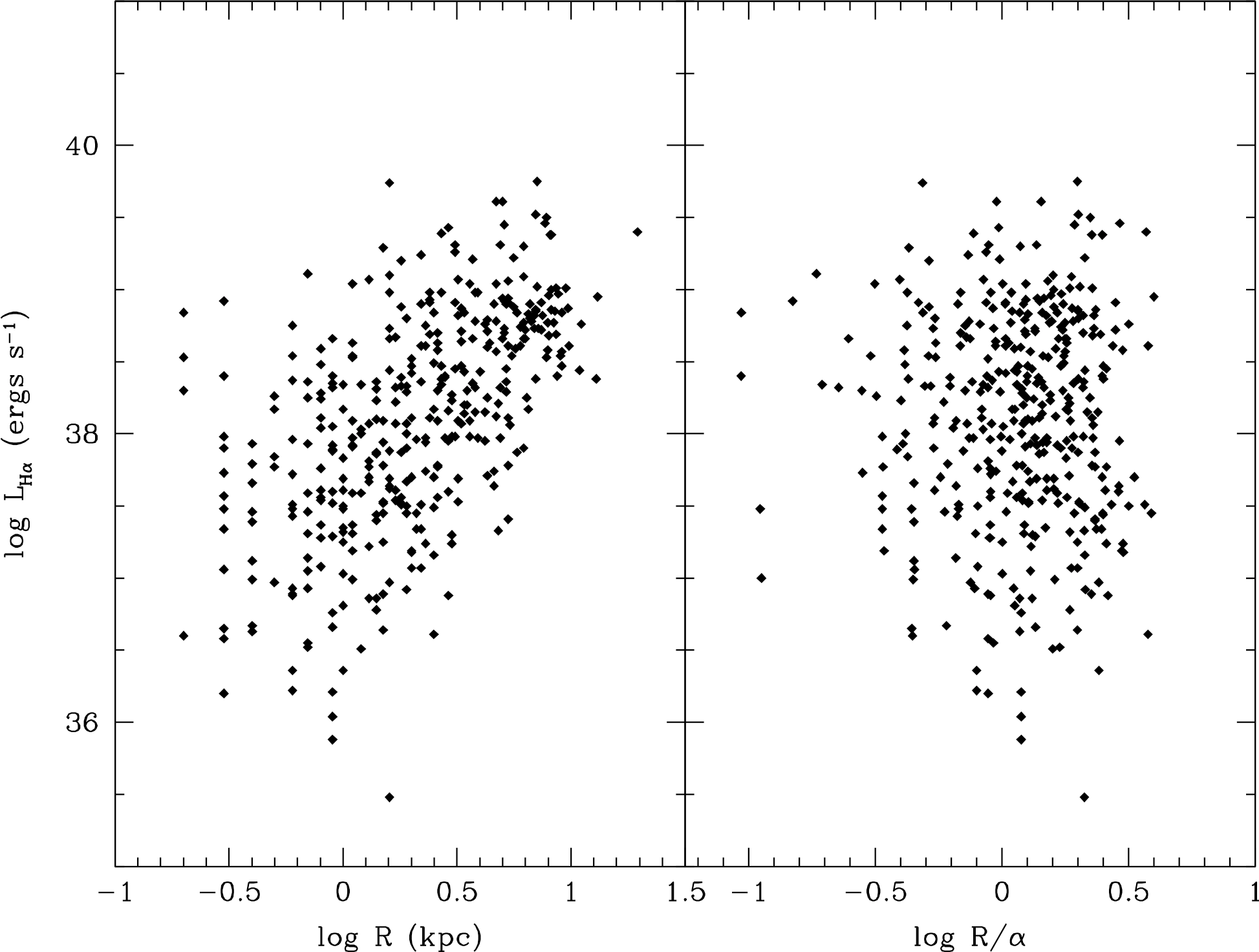}
\caption{\small HII region luminosity as a function of distance from the
galaxy center in terms of absolute kpc's and normalized scale lengths.  The
artificial relation between kpc and luminosity in the left panel is due to
the fact that the largest galaxies have the brightest HII regions.  When
galactic distance is normalized by galaxy scale length ($\alpha$), the
relationship disappears.  Since the HI densities of LSB galaxies are relatively
constant (de Blok, McGaugh \& van der Hulst 1996), this diagram simply
reflects that fact that local density drives star formation in LSB galaxies
rather than global patterns found in spirals.
}
\label{central_dist}
\end{figure}

The relationship between the H$\alpha$ luminosity of each HII region and
its distance from the galaxy center is shown in Figure \ref{central_dist}.
While the absolute distance, in kpcs, displays a trend that the
brightest HII regions are found in the outer regions (left panel), this is
an artifact of the effect that the largest (brightest) galaxies in the
sample have the brightest HII regions.  When the distance from the galaxy
core is displayed in terms of the scale length of the galaxy ($\alpha$,
from exponential fits to the $V$ frames), the relationship disappears
(right panel).  The lack of radial correlation in Figure \ref{central_dist}
is reinforced by the fact that the location of the brightest HII regions
are also independent of their distance from the galaxy center.

Our interpretation for a lack of correlation between HII region luminosity
and distance from the galaxy center is that this reflects the underlying
gas distribution in LSB galaxies.  In general, the HI gas density in LSB
galaxies is much more extended than the optical image and the density
levels are flat out to several optical scale lengths (de Blok, McGaugh \&
van der Hulst 1996).  While it is the molecular gas, not neutral hydrogen,
that drives star formation (Scoville 2012), the distribution of H$_2$ gas
in LSB galaxies is not directly known (Matthews \etal 2005) and HI serves
as a necessary proxy.  However, since the density of HI gas in LSB galaxies
is low (as are their stellar densities) and typically constant with radius
(stellar surface brightness profiles are also very shallow exponentials),
the lack of a radial trend in decreasing gas density with radius means that
star formation will be dominated by local density enhancements rather than
global processes.  And, as concluded by other studies, it is clear that the
spatial distribution of star formation in LSB galaxies differs from the
global patterns found in spirals (Bigiel \etal 2008, O'Neil, Oey \& Bothun
2007).

Presumably, the SFR will halt when the molecular gas surface density drops
below a critical value, but an estimate of where that radius occurs
requires more HI information than is available for our sample.  However,
there are numerous examples of HII regions at very low surface brightnesses
in LSB galaxies (see Figure \ref{compare_aps} for an example where
H$\alpha$ emission is found beyond 5 scale lengths).  Over 1/3 of the HII
regions in our sample occur in regions where the surface brightness is
below 25 $V$ mags arcsecs$^{-1}$ (which corresponds to less than 4
$L_{\sun}$ pc$^{-2}$) and 1/2 the HII regions have no optical signature (an
optical knot or surface brightness enhancement) even at such low surface
brightnesses (indicating a very low cluster mass).  This is an important
observation with respect to LSB galaxies as star formation has always been
assumed to be inhibited in low density environments, but not non-existent.

Star formation, as traced by H$\alpha$ is loosely correlated with optical
surface brightness in LSB galaxies, in the sense that for HII regions
without detectable optical knots there is the trend that the brightest HII
regions are located in regions of the galaxy with higher surface
brightness.  However, the trend is by no means exact and there exist many
examples of strong HII regions in areas of very low stellar density.
Gravitational instability models suggest a threshold for star formation
where the gas density falls below a critical value (Kennicutt 1998) and
star formation efficiency in HSB galaxies generally follows stellar
densities more strongly than gas densities (Leroy \etal 2008).  But, star
formation in the low surface brightness regions of our sample suggests some
other method allows the formation of the cold phase of the ISM without the
gravitational pull from stellar mass (see also Thornley, Braine \& Gardan
2006) and that gravitational instability from stellar density does not play
a dominant role.

HII regions tend to avoid the cores of LSB galaxies, as can be seen in
scale length panel of Figure \ref{central_dist}.  While the central peak
of stellar luminosity in LSB galaxies is ill-defined, due to their
irregular morphology, their outer isophotes are usually fairly regular and
can be used to define a center of stellar mass.  The fact that HII regions
tend to be found in regions outside the core may simply reflect the lumpy
distribution of stars and gas in LSB's (Pildis, Schombert \& Eder 1997)
rather than formation effects (i.e., spiral bulges).  LSB galaxies rarely
have the central concentrations, bulges or even AGN behavior that would
indicate present, or past, nuclear star formation that is common in many
starburst and spiral galaxies (Schombert 1998).

\section{Brightest HII Regions}

One area where completeness is not an issue is the characteristics of the
brightest HII region in each galaxy.  This region represents
the largest site of star formation in each galaxy and, presumably, the
largest concentration of ionizing O stars.  While the HII region luminosity
function predicts the number of bright HII regions in a galaxy, there is no
particular model or framework for understanding the relationship between
the luminosity/mass of the brightest region and global characteristics of a
galaxy (Leroy \etal 2008).

\begin{figure}[!ht]
\centering
\includegraphics[scale=0.8,angle=0]{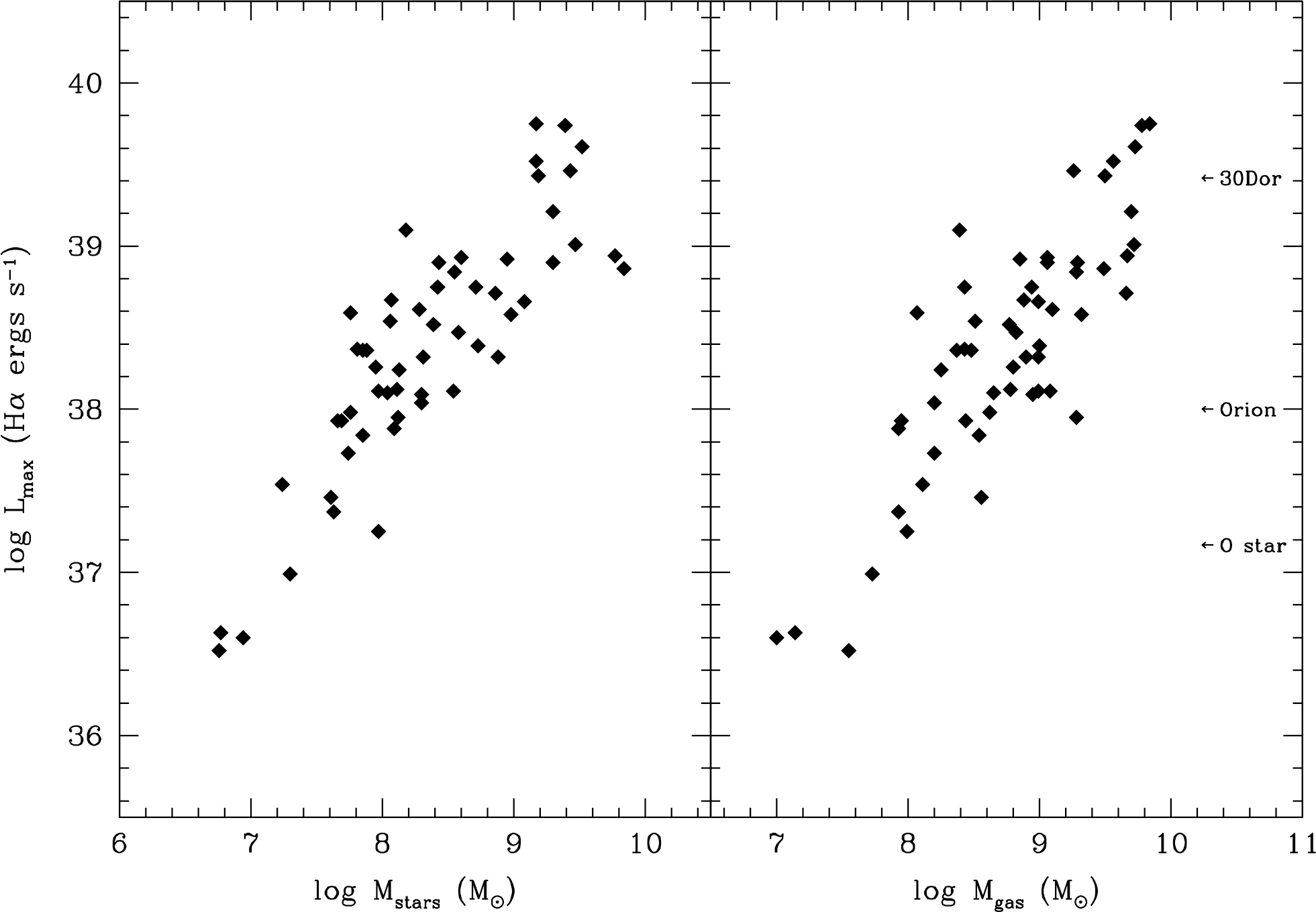}
\caption{\small The relationship between the H$\alpha$ luminosity of the
brightest HII region in each galaxy, versus the total stellar and gas mass
of the galaxy.  While a larger available gas supply would seem necessary
(if not sufficient) for a large HII region, the correlation with stellar
mass implies a longer evolutionary connection between the star formation
history of an LSB galaxy and it's current SFR.
}
\label{lmax_mass}
\end{figure}

\begin{figure}[!ht]
\centering
\includegraphics[scale=0.8,angle=0]{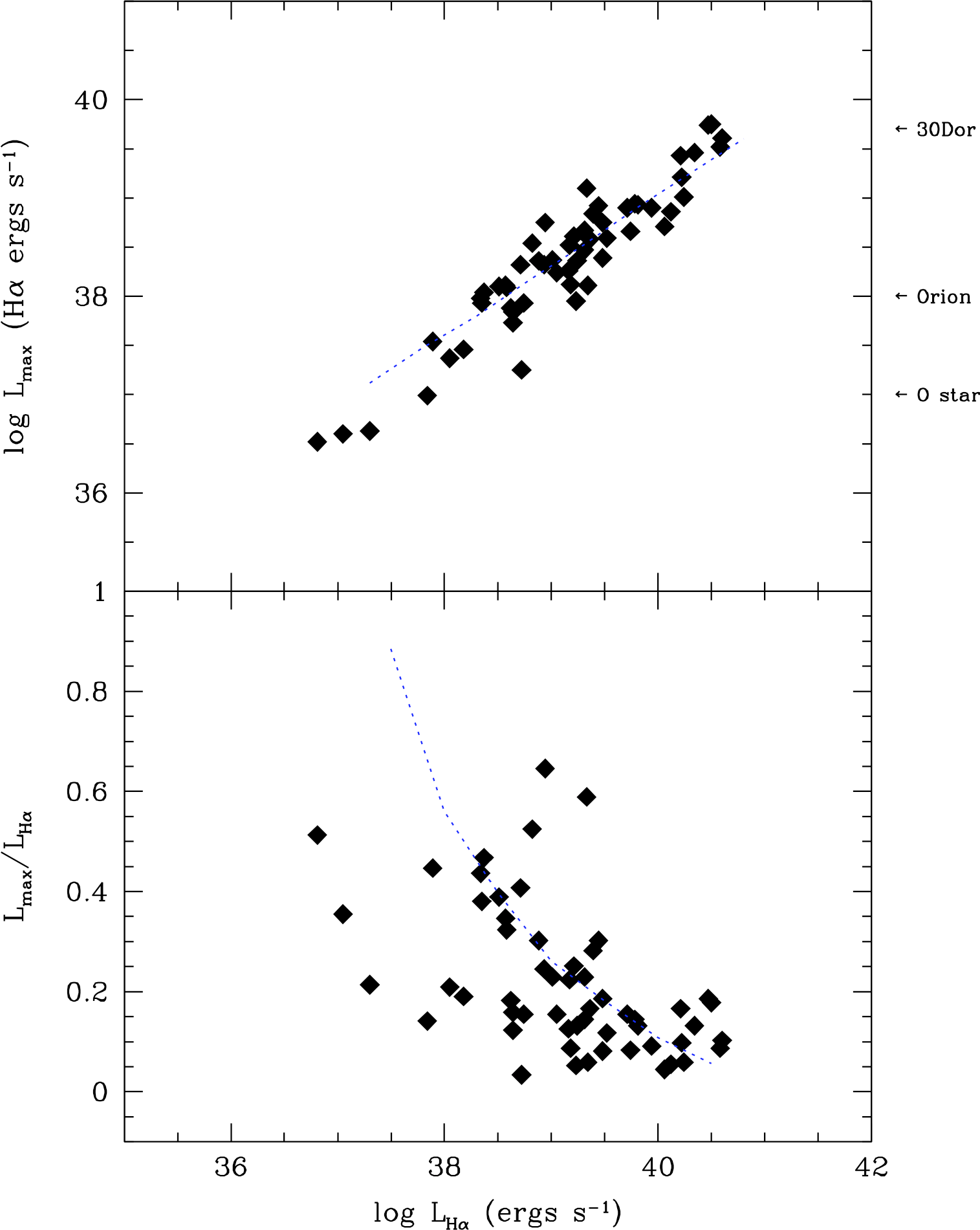}
\caption{\small The relationship between brightest HII region H$\alpha$
luminosity ($L_{max}$) and total galaxy H$\alpha$ luminosity
($L_{H\alpha}$) and the fraction of the brightest HII region to the total
galaxy flux.  The brightest HII regions (log $L_{max} >$ 39) correspond to
a cluster of several hundred O7V stars.  Yet, the fractional contribution
to the total galaxy H$\alpha$ luminosity decreases to less than 20\% for
the brightest galaxies.  The blue lines display the result of a Monte Carlo
simulation that selects HII regions from the luminosity function defined
for dwarf irregulars by Youngblood \& Hunter (1999).  There is no
indication that the HII regions in LSB galaxies display any difference from
HII regions in other irregular galaxies.
}
\label{lmax_lha}
\end{figure}

There are clear, distinct correlations between the luminosity of the
brightest HII region ($L_{max}$) and galaxy luminosity (i.e., a proxy for
stellar mass), gas mass and the total H$\alpha$ luminosity of the
galaxy.  The first two correlations are shown in Figure \ref{lmax_mass},
where stellar luminosity is converted to stellar mass following the
prescription of McGaugh \& de Blok (1997) and gas mass corrected from HI
mass for metallicity and molecular contributions.  The correlation with
total H$\alpha$ luminosity is shown in Figure \ref{lmax_lha}.

If the amount of local star formation is determined by a random process of
gas collection (e.g., cloud-cloud collisions), then the correlations with
galaxy mass would simply reflect the statistical nature of more star
formation with larger gas mass and a higher chance of building a large,
bright HII complex with more available gas.  In that scenario, the
correlations should be stronger with gas mass versus stellar mass (as the
available gas reservoir is the fuel for star formation, not stellar mass),
and the fact that there is no significant difference may signify at strong
evolutionary connection between the formation of stellar mass and the
available gas supply.  At the very least, the current SFR in an LSB galaxy
has a strong evolutionary connection with its past as defined by stellar
mass build-up, even if a significant fraction of the current SF is
occurring in low stellar density regions (perhaps future HSB regions).

The statistical nature can be understood better in terms of comparing the
total star formation rate of galaxy (as given by the total $L_{H\alpha}$)
and the luminosity of the brightest HII region ($L_{max}$).  The star
formation rates of LSB galaxies are low compared to other irregular
galaxies (Schombert, Maciel \& McGaugh 2011).  However, if the distribution
of HII region luminosities follows the same luminosity function as other
galaxies, then there should be simply a smaller number of HII regions that
can form for a given value of total $L_{H\alpha}$.  Thus, the
probability of finding a HII region of a particular luminosity decreases
with higher HII region luminosities.  

The correlation between total galaxy H$\alpha$ luminosity and the
luminosity of the brightest HII region is found in Figure \ref{lmax_lha}
(top panel), along with the ratio of the brightest HII region luminosity
and the total H$\alpha$ luminosity (bottom panel).  The brightest HII
regions correspond to approximately 200 O7V stars (Werk \etal 2008), yet as
the total SFR increases for the sample, they contribute only 10 to 20\% to
the total H$\alpha$ luminosity.  The diffuse component means that this
value will never be above 0.5 in LSB galaxies.

In order to test the idea that the properties of the observed HII regions
are simply the result of small number statistics, we constructed a simple
Monte Carlo simulation by randomly selecting HII region luminosities from
the luminosity function as given in Youngblood \& Hunter (1999) for dwarf
irregulars.  The HII region luminosities were randomly selected by their
luminosity function probability then added until the total set matched a
given $L_{H\alpha}$ value.  The luminosity of the brightest HII region was
then output.  After running 10,000 simulations per luminosity bin, the mean
brightest HII region luminosity was determined as a function of
$L_{H\alpha}$.  The results from these simulations are shown as the blue
lines in Figure \ref{lmax_lha}.

The agreement between the $L_{max}$ simulation and the data (top panel in
Figure \ref{lmax_lha}) is excellent and demonstrates that, despite previous
claims of truncated LF's in LSB galaxies (O'Neil, Oey \& Bothun 2007,
Helmboldt \etal 2005), the luminosity of the brightest HII regions is
consistent with the same pattern of HII regions for dwarf irregulars.  The
ratio of $L_{max}$ and the total H$\alpha$ luminosity is also in agreement
with the simulations, where a lack of $L_{max}/L_{H\alpha}$ near unity
simply reflects the statistical improbability of finding a single HII
region that contains all the H$\alpha$ flux of a galaxy including any
diffuse emission.

\section{Optical Colors and HII regions}

The identification of HII regions in the H$\alpha$ images allows us to use
the same apertures on the $B$ and $V$ images to extract continuum
luminosities and $B-V$ colors.  As described in \S2, we have divided the
sample of identified optical and H$\alpha$ knots into three types; 1) those
regions with H$\alpha$ emission, but no enhanced optical flux above the
mean surface brightness of the local isophote value, 2) knots with
both H$\alpha$ and optical emission and 3) knots only visible in $V$ images
without detectable H$\alpha$ emission.  These three regions would,
presumingly, correspond a low luminosity HII region (no
visible stars), a young HII region with some blowout and visible stars
(Orion type HII region) and an evolved stellar cluster or association sufficiently
old to be free of any remaining hot gas.  The regions of the first type
(no optical enhancement) are slightly redder than those HII regions with an
optical knot, but display no extra reddening compared to the regions
surrounding them.  They may, in fact, simply represent regions where the
luminosity of the underlying star cluster is small compared to the local
galaxy light, although this is a problematic interpretation due to the low
surface brightness nature of these regions.

\begin{figure}[!ht]
\centering
\includegraphics[scale=0.30,angle=0]{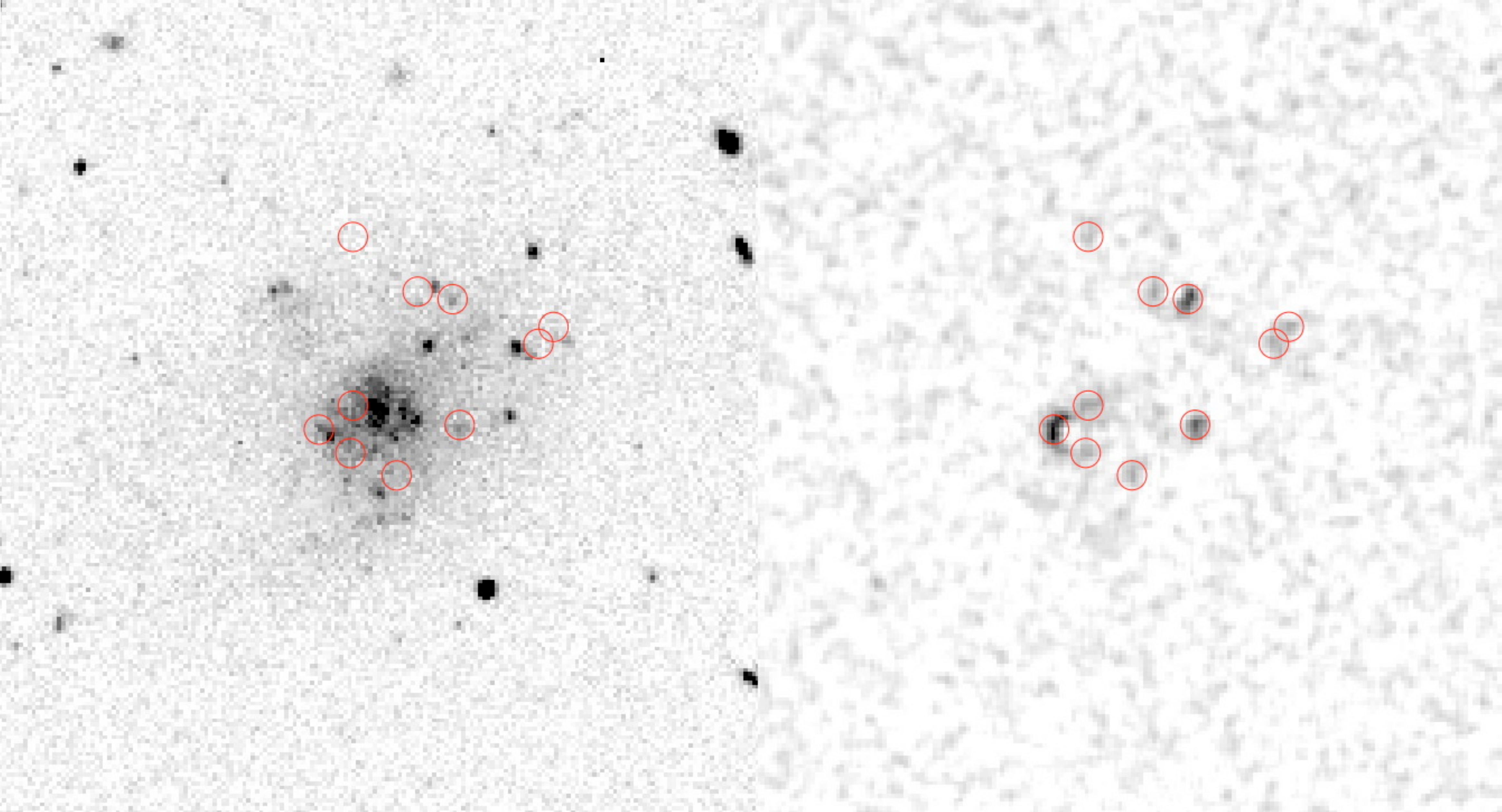}
\caption{\small The $V$ and H$\alpha$ images for LSB galaxy F608-1.  Each
frame is 150 arcsecs to a side. There are several examples of H$\alpha$
knots with no visible optical emission (the two HII regions farthest to the
right and topmost).  There are also several examples of an HII region with
a distinct optical knot in the $V$ image (e.g., the three brightest
H$\alpha$ regions).  The faintest HII regions correspond to log
$L_{H\alpha}$ between 36.2 and 36.5.  The brightest three HII regions are
log $L_{H\alpha}$ of 36.8, 36.9 and 37.0, comparable to a cluster of
stellar mass between $3\times10^3$ and $7\times10^3 M_{\sun}$ ionized by a
dozen O stars.
}
\label{compare_aps}
\end{figure}

An example of H$\alpha$ versus optical knots is shown in Figure
\ref{compare_aps}.  In this Figure, the H$\alpha$ and $V$ 
frames for F608-1 are plotted at the same scale (150 arcsecs
to a side).  The HII regions are marked in both panels by red circles, as
determined from the H$\alpha$ image.  There are several examples of
H$\alpha$ knots with no visible optical emission (the two HII regions
farthest to the right and topmost).  There are also several examples of an
HII region with a distinct optical knot in the $V$ image (e.g., the three
brightest H$\alpha$ regions).  The faintest HII regions correspond to log
$L_{H\alpha}$ between 36.2 and 36.5.  The brightest three HII regions are
log $L_{H\alpha}$ of 36.8, 36.9 and 37.0, comparable to a cluster of
stellar mass between $3\times10^3$ and $7\times10^3 M_{\sun}$ ionized by a
dozen O stars.

\begin{figure}[!ht]
\centering
\includegraphics[scale=0.8,angle=0]{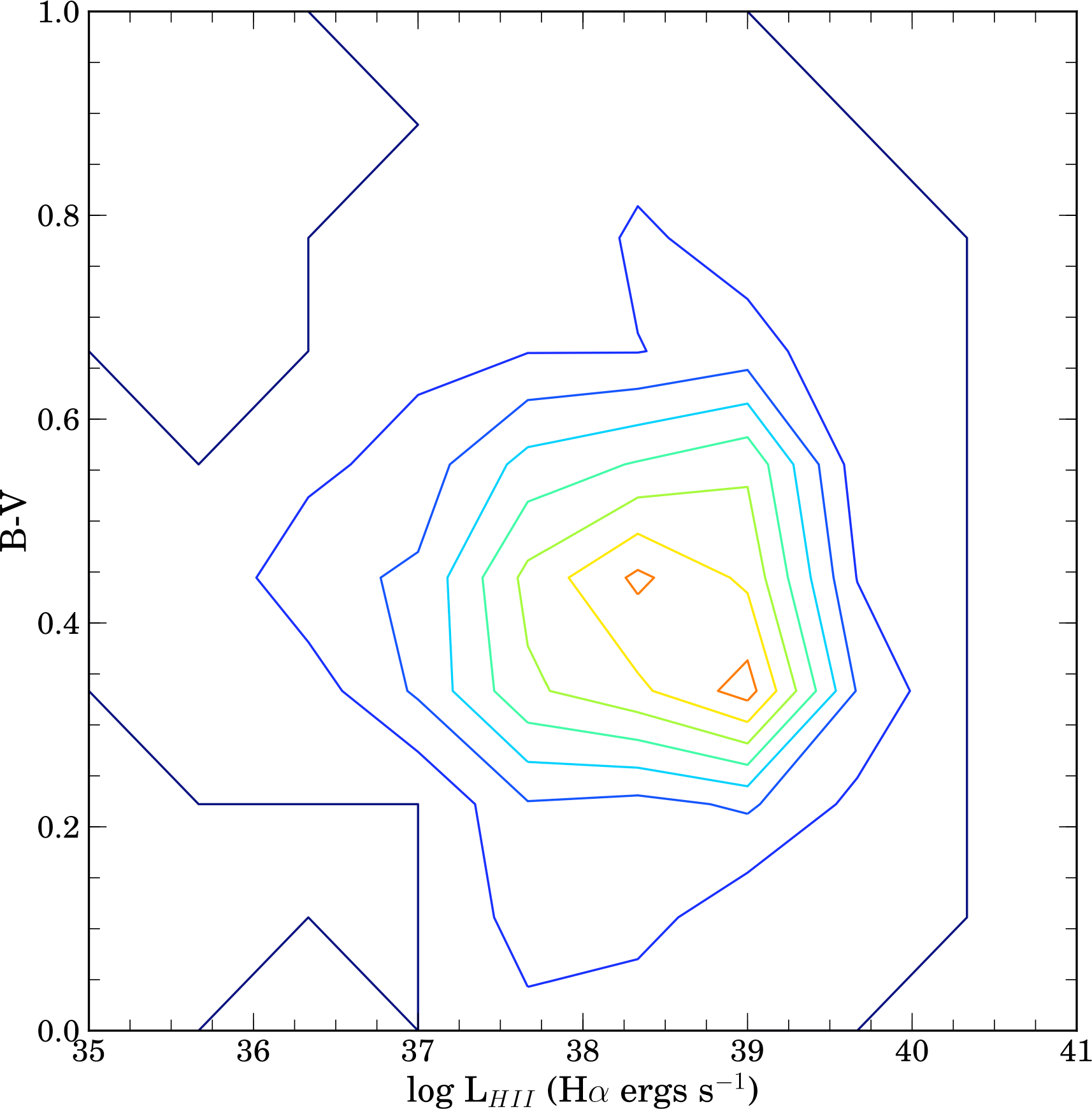}
\caption{\small A contour density plot of the H$\alpha$ luminosity versus
HII region $B-V$ color for 349 HII regions.  There is a weak trend for
bluer optical colors with increasing H$\alpha$ luminosity, consistent with
more blue ionizing stars in the brighter HII regions.  These colors are for
all HII regions with and without optical emission (star clusters) without
subtraction of the underlying galaxy light.  Regions with blue optical
enhancement will have much bluer $B-V$ colors when the surrounding galaxy
is subtracted.
}
\label{lhii_BV}
\end{figure}

For the 429 regions with H$\alpha$ emission, we have plotted their
H$\alpha$ luminosities versus their B-V colors (determined through the same
apertures as the H$\alpha$ fluxes, in Figure \ref{lhii_BV}).  No internal
extinction corrections have been applied, although gas and dust are
probably available in sufficient quantities to alter the colors.  And, more
importantly, no effort was made to subtract out the underlying galaxy light
(see below) which is necessary to compare to regions without any obvious
optical emission.

Figure \ref{lhii_BV} displays a very weak trend for bluer optical colors
with increasing H$\alpha$ luminosity.  This trend is as expected with
greater H$\alpha$ flux implying a larger number of ionizing O stars per HII
region and, therefore, greater blue flux (see Caldwell \etal 1991).
However, the poor relationship only emphasizes the rich color structure that
is found in LSB galaxies where star forming regions are often associated
with blue shells and filaments and color features uncorrelated with star
forming regions (to be studied in a later paper).  It is worth noting that
the color-H$\alpha$ trend is not as blue as HII regions in early-type
spirals (Caldwell \etal 1991).  In that sample, HII regions with log
$L_{H\alpha} = 38.5$ have $B-V$ colors less than zero.  Many of the regions
with optical emission have much bluer colors (see below) and the colors for
low luminosity HII regions are correlated with the nearby galaxy colors. We
anticipate that the underlying colors will be less than $B-V=0.0$ once the galaxy
light is subtracted.

\begin{figure}[!ht]
\centering
\includegraphics[scale=0.8,angle=0]{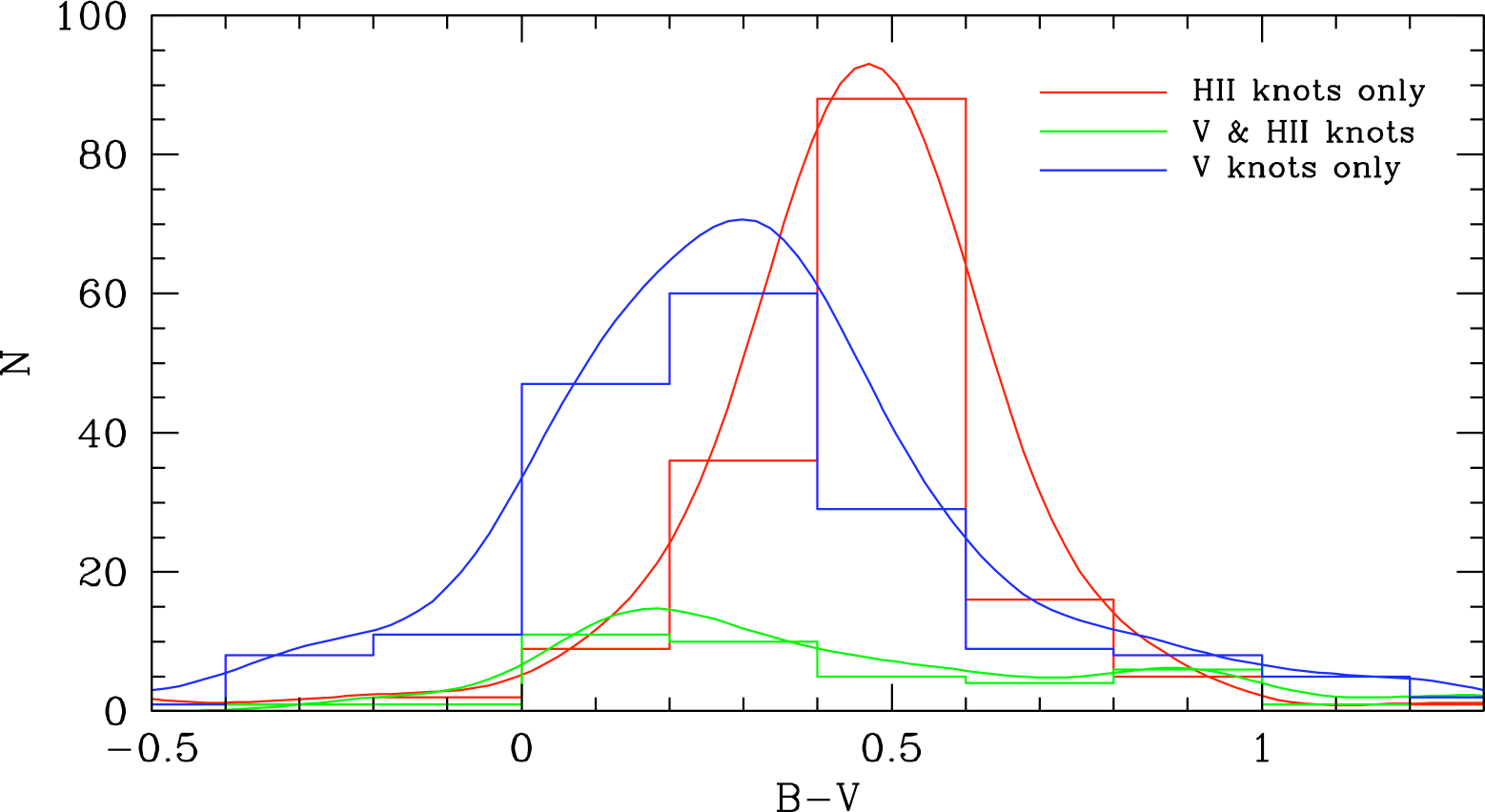}
\caption{\small Normalized absolute $V$ luminosity and color histogram for
all 318 knots with $B-V$ colors.  HII knots are regions with only H$\alpha$
emission and no visible optical enhancement above the local isophote.  $V$
and HII knots are regions with an distinct knot in both the H$\alpha$ and
$V$ images (typically brighter in H$\alpha$ than sole H$\alpha$ knots).
$V$ knots have optical emission but no detectable H$\alpha$ emission.
}
\label{color_hist}
\end{figure}

A comparative histogram of $B-V$ colors within the various H$\alpha$ and
$V$ knots is shown in Figure \ref{color_hist}.  These colors were
calculated by subtracting the local galaxy isophote from those HII regions
with optical knots, leaving only the luminosity above the underlying galaxy
luminosity density.  For HII regions without optical knots, the local
galaxy color is used.  This technique does not bias the calculated color
for the optical knots, but it was unsurprising to find the majority of them
have $B-V$ colors bluer then the local galaxy color as was noted in the
two color maps from Paper I.

Here, the reddest colors are found for the H$\alpha$ knots without any
optical signature.  It should be noted that these H$\alpha$ only knots also
typically have the lowest H$\alpha$ luminosities.  In other words, these are
regions that are ionized by a single or a very small number of O or B stars.
Their mean $B-V$ color is 0.45, which basically confirms that these regions
have little effect on the surface brightness or local color as these values
conform to the mean total color of LSB galaxies.  The underlying stellar
association lack sufficient luminosity to alter the galaxy's isophotes and
colors, even at these low surface brightness regimes ($L_{\sun}$ pc$^{-2}$
= 1 to 4).

Regions which display HII emission and an optical enhancement tend to be
bluer than sole H$\alpha$ knots (mean $B-V$ = 0.25) and are also brighter
in H$\alpha$ luminosity with values that correspond to between tens to
hundreds of O stars per region.  This trend of optical detection correlated
with H$\alpha$ emission was also seen in early-type spirals by Caldwell
\etal (1991).  The bluest knots agree well with the bluest regions for
spirals ($B-V = -0.2$).  Lastly, the optical knots without H$\alpha$
emission span a full range of $B-V$ colors, although with a mean color
slightly redder than the optical knots with H$\alpha$ emission.  The
slightly redder colors probably indicates an evolutionary effect, i.e. as a
cluster ages and the ionizing stars die off, the HII region dissipates and
the cluster ages and reddens (see below).

\begin{figure}[!ht]
\centering
\includegraphics[scale=0.9,angle=0]{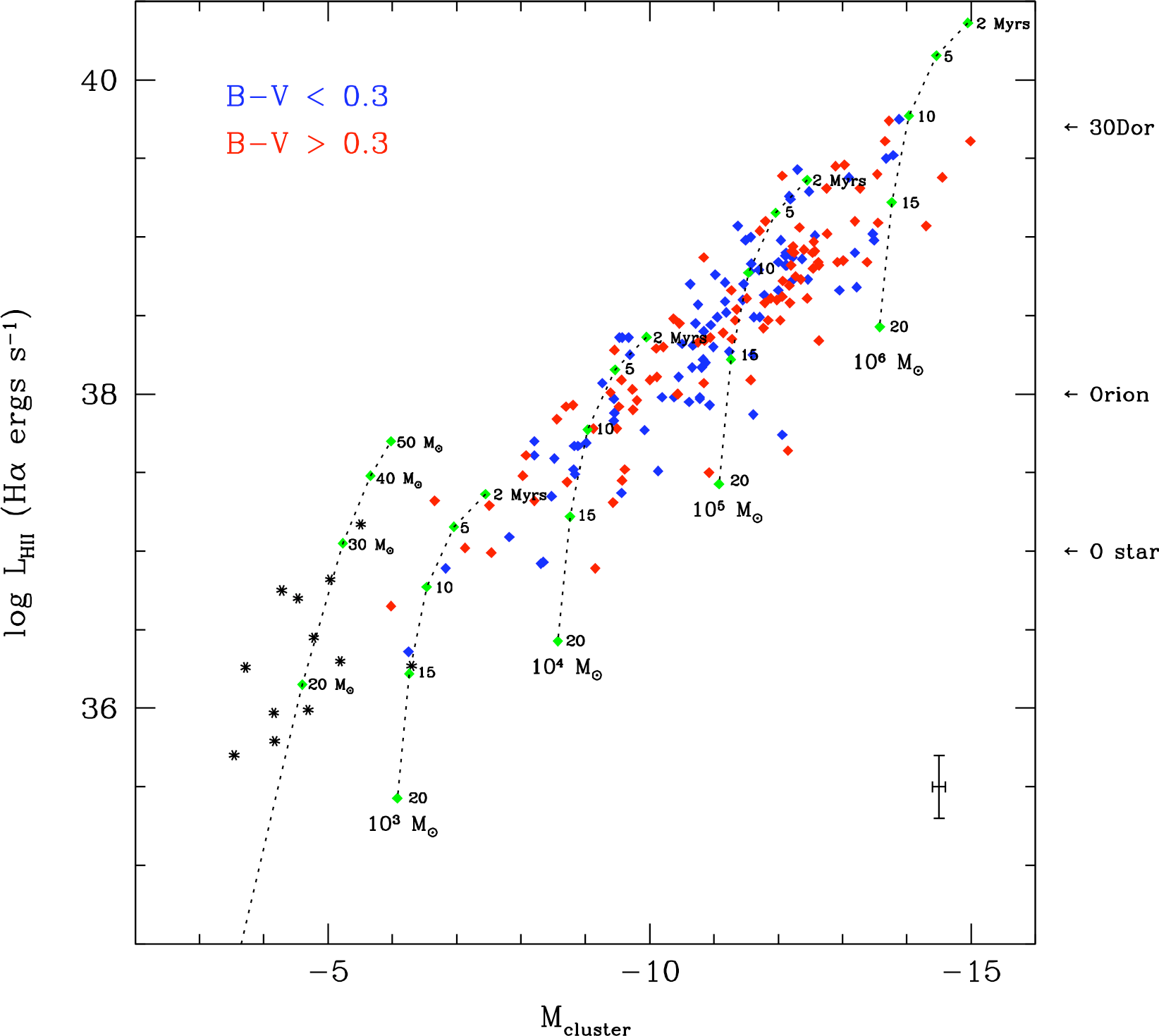}
\caption{\small For HII regions with optical signatures in the $V$ frames,
the magnitude of the underlying cluster is plotted versus the HII region
H$\alpha$ luminosity.   The correlation with brighter clusters for
increasing H$\alpha$ luminosity is clear.  Dividing the sample by a $B-V$
color of 0.3 displays a trend for the bluer clusters to be brighter in
H$\alpha$ luminosity than red clusters.  Single star HII regions from the
LMC (Zastrow, Oey \& Pellegrini 2012) are shown as black symbols.  Stellar
population models are shown as dotted tracks for cluster masses from $10^3$
to $10^6 M_{\sun}$.  Model ages are indicated in Myrs.  Typical data errors are shown in
the bottom right.
}
\label{lha_mag}
\end{figure}

To examine this evolutionary processes in greater detail, we plot in Figure
\ref{lha_mag} the absolute $V$ magnitude of the knot (presumably a
stellar association), the HII region H$\alpha$ luminosity versus
$M_{cluster}$.  The absolute $V$ magnitude of the stellar association or
cluster ranges from values of $-8$ to $-14$, which would correspond to a
range of cluster masses from open clusters to globular sized if age were
not a factor.  However, assigning a cluster mass to the $V$ magnitude is a
difficult procedure, for while the $L_{HII}$ luminosity relates the number
of ionizing O stars per region, the total number of stars (as given by the
IMF) will be extremely sensitive to the age of the HII region (Leitherer
\etal 2010).  For example, a 10 Myr $10^4 M_{\sun}$ cluster has the same
$V$ magnitude as a 100 Myr $1.5 \times 10^5 M_{\sun}$ cluster and a 500 Myr
$10^6 M_{\sun}$ cluster (Bruzual \& Charlot 2003).

We have divided the sample into red ($B-V > 0.3$) and blue ($B-V < 0.3$)
clusters.  The division of the sample by color is clear, the blue clusters
have higher $L_{HII}$ values then red clusters at constant $V$ cluster
luminosity.  The inverse interpretation, that red clusters having brighter
$V$ magnitudes at a constant H$\alpha$ value is opposite to what one would
expect from spectroevolutionary models where an aging cluster will redden
by 0.3 in $B-V$ over 500 Myrs, but the luminosity of the underlying cluster
will have decreased by 3 magnitudes.  A more plausible scenario is that age
is the defining factor in the difference between red and blue clusters in
Figure \ref{lha_mag}.  The blue clusters are younger and have more ionizing
stars per unit cluster mass producing higher H$\alpha$ luminosities.  Over
100 Myrs, the number of ionizing stars decreases by a factor of 3
(Werk \etal 2008) while the $B-V$ color has reddened by 0.2.  This is
consistent with the trend seen in Figure \ref{lha_mag}.

In order to test this hypothesis, we have constructed a series of stellar
population models taking the population colors and luminosities from
Bruzual \& Charlot (2003) for low metallicity ([Fe/H] = $-$0.4) tracks.
Starting with a given stellar mass, we apply the IMF from Kroupa \etal
(2011) to determine the number of stars with ionizing photons.  We then
apply the ionization Q curves from Martins \etal (2005) to determine the H$\alpha$
luminosity of the cluster as a function of age.  Each zero age model is
then aged used a standard stellar lifetime as a function of mass, the Q
values are recalculated and new cluster luminosities are determined.  The
resulting tracks are shown in Figure \ref{lha_mag}.

As can be seen in Figure \ref{lha_mag}, the star forming regions in LSB
galaxies range in stellar mass from globular cluster sized ($10^6
M_{\sun}$), such as 30 Doradus, to small associations ($10^3 M_{\sun}$),
such as the California nebula and the Taurus cloud in our own Galaxy.  In
addition, HII regions vary in age from 2 to 15 Myrs, although a
majority of the detected HII regions have ages between 10 and 15 Myrs.  We
note that the position of the model tracks with respect to H$\alpha$
luminosity are extremely sensitive to the shape of the upper end of the
IMF.  However, the top edge of our sample agrees well with the zero age
line from our models, indicating that the upper end of the Kroupa IMF
appears to closely represents the IMF in LSB galaxies.

Low luminosity HII regions, lacking any optical signature, would presumably
fall to the bottom left of this diagram.  For comparison, we have plotted
the data from Zastrow, Oey \& Pellegrini (2013) for single O or B clusters in
the LMC (black symbols in Figure \ref{lha_mag}).  Also shown are single
star ionization curves for single star mass of 10 to 50 $M_{\sun}$.  HII
regions with log $L_{H\alpha}$ less than 36.5 would fall in this region,
and have visual luminosities and mean surface brightnesses below detection
levels (a $10^3 M_{\sun}$ cluster within a 100pc pixel would only increase
the surface brightness of that pixel by 1\%).

\section{Conclusions}

LSB galaxies typically have low total SFR's and, thus, fewer HII regions to
study compared to spirals and irregulars.  We have attempted
to overcome this deficiency by observing a larger sample over a greater
volume of the local Universe.  Our sample of 54 LSB galaxies produced 429
HII regions for study, most having sufficient S/N in their optical images
to compare broadband luminosities and colors.  Four galaxies in our sample
were undetected in H$\alpha$ and have the lowest gas mass fraction of the
sample, suggesting their lower gas supply is responsible for their lack of
star formation.  

We summarize our results as the following:

\begin{description}

\item{(1)} LSB galaxies typically have fewer HII regions per galaxy than
other irregular galaxies; however, LSB galaxies have a full range of HII
region sizes from complexes that encompass regions powered by a single O or B
star (log $L_{H\alpha} < 36.5$) to 30 Doradus sized complexes with log
$L_{H\alpha} > 40$.  The correlation between HII region luminosity and size
is well defined with a slope of two, indicating that we are observing all of
the photons from the ionized gas.

\item{(2)} LSB galaxies have a wide range in the fraction of HII region's
contribution to the total $L_{H\alpha}$ luminosity from 10 to 90\%.  The
fraction having no correlation with galaxy baryon mass.

\item{(3)} There is no correlation between the HII region luminosity and
spatial position in a galaxy.  The brightest HII regions do not
preferentially appear at any particular radius as normalized by disk scale
length.

\item{(4)} Roughly 1/2 of the HII regions have a distinct optical
enhancement above the surrounding isophote.  This is interpreted to be
stellar mass produced by the star formation event (which is confirmed by
their bluer colors compared to surrounding galaxy color).  HII regions
without enhancement are still, loosely, associated with local stellar
density (i.e., surface brightness) in proportion to their $L_{H\alpha}$.
However, there are numerous examples of bright HII regions in faint galaxy
regions.  

\item{(5)} The luminosity of the brightest HII region in each galaxy is
correlated with the galaxy's stellar mass, gas mass and total star
formation rate.  Monte Carlo simulations confirm that these correlations
are replicated by an underlying HII region luminosity function that matches
that for star forming irregulars.  In other words, there is no evidence
that the distribution of HII regions luminosities in LSB galaxies differ from
that of star forming HSB galaxies, and the underlying star formation
mechanisms appear to be the same.

\item{(6)} As observed in spiral galaxies, there is a weak correlation
between the color of a HII region and its H$\alpha$ luminosity.  And, while
regions with H$\alpha$ emission are bluer with increasing H$\alpha$
luminosity, there are blue regions in a LSB galaxy without H$\alpha$
emission.

\item{(7)} Comparison with stellar population models indicates that the HII
regions in LSB galaxies range in mass from a few $10^3 M_{\sun}$ to
globular cluster sized systems.  Their ages are consistent with clusters
between from 2 to 15 Myrs old.  The faintest HII regions are also similar
to single O or B star associations seen in the LMC.  Thus, star formation in
LSB galaxies covers the full range of stellar cluster mass and age.

\end{description}

The hope in studying LSB galaxies was to reveal, perhaps, a new realm of
star formation processes or conditions.  Where the class of LSB galaxies
differ from HSB galaxies in terms of their bluer colors, lower stellar
densities and higher gas fractions; however, there is nothing particularly
unusual about the individual sites of star formation under more detailed
examination.  The local process of star formation, cluster size and mass,
IMF and gas physics, all are consistent with the style of star formation
found in HII regions in spirals and irregulars.  With respect to their
global properties, the HII regions in LSB galaxies are more similar to
other irregular galaxies, again reflecting the sporadic distribution of gas
over coherent kinematic processes (i.e., spiral patterns).

\acknowledgements We gratefully acknowledge KPNO/NOAO for the telescope
time to complete this project.  Software for this project was developed
under NASA's AIRS and ADP Programs.

\end{document}